\documentclass[conference]{IEEEtran}
\IEEEoverridecommandlockouts

\usepackage{cite}
\usepackage{subcaption}
\usepackage{amsmath,amssymb,amsfonts}
\usepackage{algorithmic}
\usepackage{graphicx}
\usepackage{textcomp}
\usepackage{xcolor}
\def\BibTeX{{\rm B\kern-.05em{\sc i\kern-.025em b}\kern-.08em
    T\kern-.1667em\lower.7ex\hbox{E}\kern-.125emX}}
\begin{document}

\title{MultModLM: A multi-modal benchmark for Large-Language Model based hardware schematic generation}

\author{\IEEEauthorblockN{1\textsuperscript{st} Dhruv Kulkarni}
\IEEEauthorblockA{\textit{DoCSE} \\
\textit{SVNIT}\\
Surat, India \\
dhruvkulkarni22@gmail.com}
\and
\IEEEauthorblockN{2\textsuperscript{nd} Sai Manoj Pudukotai Dinkarrao}
\IEEEauthorblockA{\textit{DoECE} \\
\textit{George Mason University}\\
Fairfax, India \\
spudukot@gmu.edu}

}

\maketitle

\begin{abstract}
Recently, Large Language models (LLMs) find application in several fields. This extends to hardware definition and synthesis. However, most works at the intersection of LLMs and hardware generation focus on text-based tasks, creating a gap for multi-modal LLMs for RTL design.
In this work, we introduce \textbf{MultModLM}, a benchmark for evaluating LLMs on the task of generating hardware schematics from RTL (Register Transfer Level) descriptions. The dataset consists of 99 diverse RTL modules spanning arithmetic, control, and state-based designs. To address the challenges of non-unique schematic representations, we propose a multi-stage evaluation framework combining rubric-based scoring, self-evaluation, cross-model assessment, blind evaluation, and human validation to enable exhaustive evaluation.

Through experiments on state-of-the-art LLMs, we observe that while models can generate visually interpretable schematics, their functional correctness remains constrained. Furthermore, we find that LLM-based evaluators exhibit near-zero agreement with human raters, revealing, as a key finding, that LLM-as-a-judge paradigms are unreliable in structurally precise domains. These findings suggest that reliable evaluation of multi-modal hardware outputs remains an open challenge, motivating the need for more robust and domain-aware evaluation methodologies, as well as tools for structural evaluation, so as to enable formal equivalence checkers.
\end{abstract}


\section{Introduction}

Large language models (LLMs) find increased utilization in varied 
fields, including hardware development and simulation \cite{LLMSurvey}. 
In various stages and sub-stages of chip design, large language models 
are being utilized to enhance performance and security measures 
\cite{LLMSurvey}. Despite this progress, existing works have primarily 
focused on text-based tasks, such as code generation (\cite{rtllm}, 
\cite{hdlbits}), code debugging (\cite{autochip}), and evaluation 
(\cite{verilogeval}). This creates a crucial gap, since no benchmark 
exists for evaluating LLMs on multimodal hardware tasks, in particular, 
the generation of hardware schematics from RTL descriptions. 
Given the importance of hardware schematics in education as well as industry, it is crucial to attempt to ensure schematic generation is a lightweight and simple process, since it would be applicable in lowering the barrier to hardware design as well as utilizing this attempt to evaluate the capability of LLMs within the structural domain.

Traditionally, generating schematics from RTL requires expensive 
proprietary EDA tools such as Synopsys or Cadence, which demand 
significant domain expertise and are inaccessible to many practitioners 
and researchers owing to costs. Attempting to extend existing tools like Vivado to full image or schematic generation will require extensive licensing, strong domain expertise, and at times proprietary information. The goal of this work is to attempt to create lightweight schematic generators using LLMs, as LLMs have already taken part in the Hardware Development Life cycle.

To address and study the capability or inefficiency of LLMs in hardware schematic generation requires the details of the instances where they fail as well as the extent. For visual based RTL outputs, to our knowledge, no such benchmark exists. This absence makes it 
impossible to systematically measure LLM performance, 
compare models, or identify failure modes in this domain. Therefore, this is a core bottleneck preventing further progress. To address this bottleneck, we introduce MultModLM, a novel benchmark dataset for RTL to schematic generation.

The generation of schematic from RTLs task is fundamentally different from code generation, as it requires 
models to capture structural, functional, and visual aspects of digital 
circuits simultaneously. The importance of text-to-visual LLM pipelines 
and their cross-domain validity has been demonstrated in works such as 
\cite{sciimage}, further motivating exploration in the hardware domain.

Evaluating LLM-generated schematics is non-trivial and 
presents unique challenges. First, schematic representations are not 
unique:  multiple structurally different schematics can be functionally 
equivalent, meaning there is no single ground truth. Second, unlike 
code, visual outputs cannot be directly simulated or formally verified. 

 We define hardware correctness along 
three axes: (1) functional consistency with the RTL description, 
(2) structural fidelity capturing datapath and control organization, 
and (3) schematic validity and legibility, including timing semantics 
and signal integrity.

The MultModLM benchmark dataset consists of 99 RTL modules spanning arithmetic 
units, finite state machines, and control logic. We propose a 
multi-stage evaluation pipeline combining rubric-based scoring, 
self-evaluation, cross-model assessment, blind evaluation, and human 
validation.  The 99-module dataset, despite being moderate in size, covers the primary circuit categories encountered in RTL design. The benchmark is not intended for deep model training but for evaluation, in particular for probing LLM capability on a structured and diverse set of circuits. Future work would delve into extending the dataset for fine-tuning purposes as well as a deep reset purpose.
Our central finding is that LLM-based evaluators exhibit 
near-zero agreement with human raters (Cohen's $\kappa \approx 0$), 
revealing a fundamental limitation of LLM-as-a-judge in this domain 
and motivating the need for more robust, domain-aware evaluation 
methodologies.

The main contributions of this work are as follows:

\begin{itemize}
    \item We introduce a novel multimodal benchmark for evaluating 
    RTL-to-schematic generation using LLMs.
    \item We propose a structured evaluation framework combining 
    rubric-based scoring, cross-model assessment, and human evaluation.
    \item We provide empirical evidence that LLM-based evaluators 
    exhibit near-zero agreement with human judgment in hardware 
    schematic evaluation, highlighting a fundamental limitation of 
    LLM-as-a-judge in structurally precise domains.
\end{itemize}






\section{Benchmark Dataset}
\subsection{Dataset Construction}

In a typical EDA flow, schematics serve as the bridge between 
high-level RTL descriptions and physical layout. They provide a 
human-interpretable visual representation of the circuit's logical 
structure, enabling design verification, debugging, and communication 
between engineers before committing to synthesis. Traditionally, 
schematics are produced by expensive proprietary tools such as 
Synopsys or Cadence, requiring significant domain expertise. 
Automating this step via LLMs is therefore a practically relevant 
and an underexplored problem — yet no benchmark exists to evaluate 
whether LLMs can do this reliably.

To fill this gap, we construct \textbf{MultModLM}, a dataset of 
99 RTL modules that serve as prompts for LLM-based schematic 
generation. To be explicit about the division of labor: we provide 
the RTL descriptions; the LLM is tasked with generating the 
corresponding hardware schematics. The RTL modules were sourced 
from open-source repositories such as OpenRISC \cite{openrisc_or1200} 
and human generation. No ground-truth schematics are provided, 
as schematic representations are not unique, since different synthesis 
strategies can produce structurally different yet functionally 
equivalent circuits. Correctness is instead assessed through the 
multi-stage evaluation framework described in Section III.

The RTL designs span a range of algorithmic and structural patterns 
commonly encountered in digital logic design. The majority of 
modules implement numeric property checking circuits, including 
tasks such as Armstrong number detection, palindrome detection, 
and other number-based predicates. These designs involve arithmetic 
operations, conditional logic, and iterative computations over 
digit representations, making them particularly challenging for 
LLMs to construct schematics for, thus stress-testing the 
benchmark effectively on moderate complexity. In addition, the dataset includes finite state machine (FSM)-based detectors, arithmetic datapath components, 
simple counters, and control logic modules. This diversity ensures 
coverage of both combinational datapath logic and sequential 
state-based designs, which are common in RTL implementations. 
Each design is provided as a standalone Verilog module, with 
varying levels of complexity, ranging from small combinational 
circuits to multi-stage sequential designs.

\subsection{Dataset Characteristics}
The category-wise breakdown of the dataset is given in Table \ref{tab:dataset}. 
We design our dataset to emphasize algorithmically complex circuits, such as numeric property checkers, which require iterative computation and conditional logic, alongside finite-state machine (FSM) based designs that stress sequential behavior and control dependencies.

\begin{table}[h]
\centering
\caption{RTL line counts.}
\label{tab:rtl_complexity}

\begin{tabular}{lcc}
\hline
\textbf{RTL Lines} & \textbf{Count} & \textbf{Avg Lines} \\
\hline
1--20 & 30 & 15.6 \\
21--40 & 36 & 27.4 \\
41--60 & 18 & 45.3 \\
61--80 & 6 & 68.5 \\
80+ & 9 & 430.7 \\
\hline
\end{tabular}

\end{table}

\vspace{0.5em}

\begin{table}[t]
\centering
\begin{tabular}{lc}
\hline
Category & Count \\
\hline
Counters & 2 \\
Arithmetic & 5 \\
FSM Detectors & 13 \\
Numeric property checkers & 76 \\
Control & 3 \\
\hline
Total & 99 \\
\hline
\end{tabular}
\caption{Functional composition of the benchmark dataset.}
\label{tab:dataset}
\end{table}

\vspace{0.5em}
\subsection{Evaluation Principle}

Given the variable nature of schematic representations, depending on different optimization techniques, different schematics can be produced. Therefore, we do not insert a shared ground truth as the correct schematic. Instead, correctness is determined through 
functional equivalence between the generated schematic and the 
original RTL description, assessed through exhaustive rubrics and 
multi-stage evaluation. This paves a road for robust evaluation, with different stages compensating for different failure modes of each other.

\subsubsection{Functional Correctness Approximation}

Complete Formal equivalence checking between generated schematics and RTL 
descriptions are not performed, as this would require converting 
visual outputs back into executable representations, either graph or other structures. This is a process 
that is non-trivial, computationally expensive, and outside the scope of this work. Therefore, we approximate functional correctness through a combination of 
rubric-based evaluation and grounded comparison with the original 
RTL code. The rubric explicitly captures key aspects such as 
datapath behavior, control logic, and timing semantics, enabling 
detection of inconsistencies or incorrect transformations. 
Grounded cross-evaluation further ensures that the generated 
schematic is assessed relative to the RTL specifications, reducing 
the likelihood of structurally valid but functionally incorrect 
outputs.

\subsubsection{Limitations and Justification}

We acknowledge that this approach does not guarantee formal 
correctness, for no rubric-based method can fully compensate a lack of formal equivalence checking. Specifically, two key limitations 
exist: (1) Rubric scoring depends on the evaluator's ability to 
correctly interpret the schematic, which is itself unreliable 
for LLM-based evaluators as our Kappa results demonstrate; and 
(2) visual schematics may appear structurally plausible while 
containing subtle functional errors that neither the rubric-based 
nor can cross-model evaluation reliably detect.

However, this approximation is both necessary and justified for 
two reasons. First, exact equivalence checking for visual outputs 
is an unsolved problem. To our knowledge, no existing tool can formally verify 
a generated schematic image against RTL without extensive 
preprocessing. Second, the goal of this benchmark is not to 
certify correctness, but to evaluate the relative capability 
of LLMs in schematic generation and to expose the limitations 
of LLM-as-a-judge in this domain. A structured approximation 
is therefore the appropriate methodology for a benchmark of 
this nature, and is consistent with how correctness is 
approximated in other multi-modal generation benchmarks 
\cite{sciimage}. Establishing a fully automated and formally 
grounded evaluation method remains an important direction for 
future work.
\section{Experimental Methodology}

\begin{figure*}[!t]
\centering
\includegraphics[width=\textwidth]{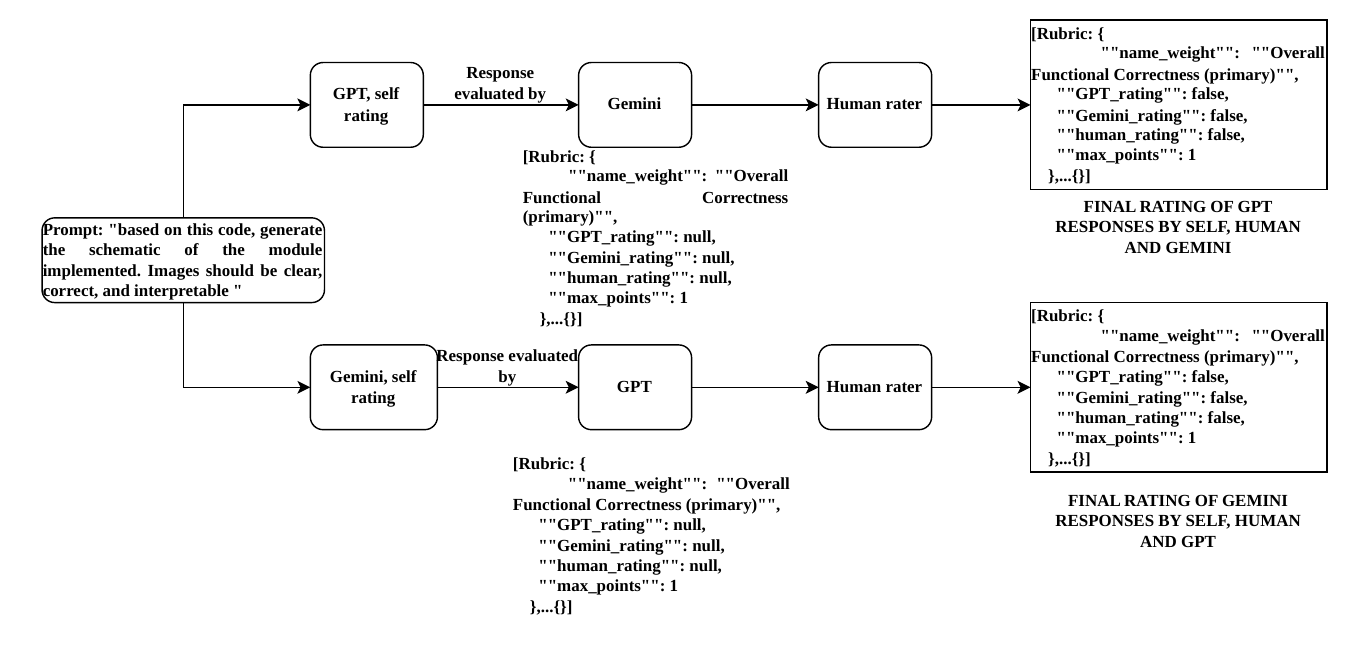}
\caption{Evaluation pipeline for comparing GPT and Gemini schematic generation.}
\label{fig:pipeline}
\end{figure*}

As no single evaluation method is sufficient, considering formal verification 
is infeasible for visual outputs, self-evaluation is prone to bias, 
and human evaluation is expensive and not scalable, we adopt a 
multi-stage pipeline where agreement across independent evaluators 
approximates correctness. Each stage addresses a different failure 
mode of the others, as illustrated in Figure \ref{fig:pipeline}.
Our methodology is distinguished from naive schematic generation in three ways:
1) we evaluate LLMs as a zero shot schematic generator without any tool access
2) we propose a rubric-based evaluation framework applicable where formal verification is unavailable 3)Because we employ LLMs as a judge, it paves the way to study the capability of LLMs as evaluators in structural domains.

\subsection{Evaluation Pipeline}

\subsubsection{Prompt Structure}
Each RTL module is provided to the target model with a prompt of 
the form: \textit{"based on this code, generate the schematic of 
the module implemented. Images should be clear, correct, and 
interpretable [RTL\_code]"}. We observed that, in some cases models instead of creating new images, either imported existing images from the internet, or started explaining the code without generating any image at all. In cases where models attempted to 
import schematics from external websites or produce textual 
explanations instead of images, an extended prompt was used with 
explicit constraints: \textit{"I ONLY WANT A BRAND NEW PNG IMAGE, 
DON'T EXPLAIN UNLESS ASKED, DO NOT USE CONTENT FROM OTHER WEBSITES"}. 
This restriction was necessary to ensure the model's output was 
original and evaluable.

\subsubsection{Self-Evaluation}
After schematic generation, the target model is asked to evaluate 
its own output within the same conversation context, using the 
predefined rubric. This stage establishes a baseline for 
self-assessment and is later compared against human and cross-model 
ratings to quantify self-assessment bias. The prompt template is: 
\textit{"Rate this response based on this rubric, fill only 
[Model]\_rating values as either true or false in boolean, leaving 
everything else as is."}

\subsubsection{Grounded Cross-Evaluation}
Following self-evaluation, the schematic generated by one model 
(the target model) is evaluated by the other model (the secondary 
model), with the original RTL code provided as contextual grounding. 
For example, when GPT generates the schematic, Gemini evaluates it 
against the RTL description, and vice versa. This cross-model 
design reduces self-assessment bias, a model rating its own output 
is more likely to be lenient than strict, and  the evaluation approximates a more objective 
assessment, analogous to inter-rater reliability in human studies. 
This is the primary stage for assessing functional correlation and 
correctness. The prompt template is: \textit{"Compare the schematic 
against the RTL. Fill the GPT/Gemini rating field as Boolean True 
or False. Do not assume correctness. Base answers strictly on 
logical equivalence."}

\subsubsection{Blind Cross-Evaluation}
In this stage, the evaluating model receives only the generated 
schematic image, without access to the original RTL code. Blind evaluation intentionally removes 
RTL access to measure schematic self-consistency and 
visual coherence independently.  Blind evaluation is not intended to 
verify functional correctness against the RTL, which is 
the role of grounded cross-evaluation.

\subsubsection{Human Evaluation}
A human evaluation step is conducted to provide an independent 
ground truth assessment. Human evaluators perform structured 
verification using the predefined rubric, judging across all defined criterias. Human evaluation in 
this work was conducted on circuits of moderate complexity. We 
acknowledge that as schematic complexity scales, (example: 
for multi-module or hierarchical designs) human validation 
becomes exponentially time-intensive and error-prone. This 
represents a central open challenge for the field, motivating 
future work on automated structural verification tools that 
can serve as ground truth proxies at scale.

\subsection{Evaluation Metrics and Design Choices}

\subsubsection{Rubric Structure}
Correctness is assessed using a rubric defined as a JSON file 
with 8 criteria, each weighted as a primary criterion, since 
all are considered necessary for a fully correct schematic. For 
objectivity, each criterion is scored as Boolean True or False 
only, avoiding the subjectivity introduced by graded scales. 
Our philosophy for doing so is that hardware schematic components such as clocking, timing, and 
labelling is inherently discrete and non-gradual; graded 
evaluation would introduce ambiguity in domains that require 
logical precision. The eight criteria are:

\begin{itemize}
    \item \textbf{Structural Fidelity to Source Description or Code:}
    Whether the generated schematic preserves the overall structural 
    organization of the RTL description, including registers, 
    combinational logic blocks, and interconnections between different modules.

    \item \textbf{Correct Data Path Representation:}
    Whether arithmetic operations, signal flows, and combinational 
    logic is correctly depicted, including placement of operators 
    and routing of signals between data-processing components.

    \item \textbf{Correct Control Signal Representation :}
    Whether control logic including enable signals, select lines, FSM 
    control paths, is correctly represented and corresponds 
    logically to the RTL behavior.

    \item \textbf{Clocking and Timing Accuracy:}
    Whether clock-driven behavior is correctly shown, with 
    sequential elements properly identified as clocked components 
    and timing semantics such as \texttt{posedge clk} preserved.

    \item \textbf{Reset and Initialization Behavior Representation:}
    Whether reset signals and initialization logic are properly 
    included, with reset paths correctly connected to sequential 
    elements.

    \item \textbf{Signal and Component Label Accuracy:}
    Whether signal names and component labels are consistent with 
    the RTL description, with no hallucinated or incorrectly 
    named signals.

    \item \textbf{Visual Clarity and Interpretability:}
    Whether the schematic is clearly organized, legible, and 
    interpretable without additional explanation, with 
    non-overlapping signals and faithful component representation.

    \item \textbf{Overall Functional Correctness:}
    A holistic assessment of whether the schematic faithfully 
    captures the functional intent of the RTL description across 
    all of the above dimensions.
\end{itemize}
\begin{table}[h]
\centering

\caption{Rubric criteria mapping (C1--C8) with example ratings.}
\label{tab:rubric}
\begin{tabular}{|c|l|c|c|c|}
\hline
\textbf{ID} & \textbf{Criterion} & \textbf{GPT} & \textbf{Gemini} & \textbf{Human} \\
\hline
C1 & Overall Functional Correctness & \texttimes & \texttimes & \texttimes \\
\hline
C2 & Structural Fidelity to Source Code & \texttimes & \checkmark & \texttimes \\
\hline
C3 & Correct Data Path Representation & \texttimes & \texttimes & \texttimes \\
\hline
C4 & Correct Control Signal Representation & \texttimes & \texttimes & \texttimes \\
\hline
C5 & Clocking and Timing Accuracy & \texttimes & \checkmark & \texttimes \\
\hline
C6 & Reset and Initialization Behavior & \texttimes & \checkmark & \texttimes \\
\hline
C7 & Signal and Component Label Accuracy & \texttimes & \texttimes & \texttimes \\
\hline
C8 & Visual Clarity and Interpretability & \texttimes & \texttimes & \texttimes \\
\hline
\end{tabular}

\end{table}

\subsubsection{Evaluation Validity}
Direct functional verification of generated schematics is 
non-trivial, as visual outputs cannot be readily simulated or 
formally verified without exhaustive preprocessing. Taken together, these four stages cross-check each other's weaknesses. No single stage is treated as the final word, agreement across all of them is what decides the final correctness assessment.

\subsubsection{Comparison with Prior Works}
We compare MultModLM with previous LLM-based hardware benchmarks 
in Table \ref{tab:comparison}. To our knowledge, this is the 
first work that explores multimodal feedback for hardware 
generation using LLMs, where VO refers to Visual Output.
\begin{table*}[t]
\centering
\caption{Comparison of MultModLM with prior LLM-based hardware benchmarks.}
\label{tab:comparison}

\begin{tabular}{lcccccc}
\hline
\textbf{Work} & \textbf{Task} & \textbf{Modality} & \textbf{Output} & \textbf{Evaluation} & \textbf{Multimodal} & \textbf{VO} \\
\hline

RTLLM~\cite{rtllm} 
& NL $\rightarrow$ RTL 
& Text 
& Verilog Code 
& Functional correctness 
& $\times$ 
& $\times$ \\

VerilogEval~\cite{verilogeval} 
& NL $\rightarrow$ RTL 
& Text 
& Verilog Code 
& Testbench-based evaluation 
& $\times$ 
& $\times$ \\

HDLBits Benchmark~\cite{hdlbits} 
& Problem $\rightarrow$ RTL 
& Text 
& Verilog Code 
& Pass/fail testing 
& $\times$ 
& $\times$ \\

AutoChip~\cite{autochip} 
& RTL Debugging 
& Text 
& Corrected RTL 
& Compilation + simulation 
& $\times$ 
& $\times$ \\

\textbf{MultModLM} 
& RTL $\rightarrow$ Schematic 
& \textbf{Multimodal} 
& \textbf{Circuit Diagram (Image)} 
& \textbf{Rubric + Human + Cross-eval + $\kappa$} 
& \checkmark 
& \checkmark \\

\hline
\end{tabular}

\end{table*}
\section{Experimental results}
Experiments were carried out on the GPT 5.2 Go version, and the Gemini 3 Flash Pro version. Several metrics were used:
\subsubsection{\textbf{Cohen's Kappa}}

Cohen's $\kappa$ measures the agreement between the human evaluator and the secondary model rater.

\begin{align}
\kappa &= \frac{P_o - P_e}{1 - P_e} \\
P_o &= \frac{TP + TN}{N} \\
P_e &= 
\left(\frac{TP+FP}{N} \cdot \frac{TP+FN}{N}\right) +
\left(\frac{TN+FN}{N} \cdot \frac{TN+FP}{N}\right) \\
N &= TP + FP + TN + FN
\end{align}
Here $TP$, $FP$, $TN$, and $FN$ denote true positives, false positives, true negatives, and false negatives, respectively.
The purpose of this metric is to analyze the second part of our research problem, which is to understand the judging capabilities of LLMs-as-a-judge.
The Kappa obtained for rating was:

\begin{tabular}{c|c}
\hline
       GPT vs Human Kappa: & 0.013 \\
      Gemini vs Human Kappa: &  0.000 \\
      \hline
\end{tabular}

Both models exhibit near-zero agreement with human evaluators, indicating limited reliability as automated judges
Accuracy was measured by the ratings of the secondary model.

Accuracy was measured by the rating of the secondary model. For example, when Gemini is the target model, GPT is the secondary model for evaluating correctness.
While LLM-based evaluation introduces subjectivity, it enables scalable assessment; however, we later analyze its limitations in Section VI.
As the rubric JSON had true or false as ratings, every true rating was accorded 1 point, and every false rating was accorded false.
Accuracy was considered on the basis of the number of points scored, where for each prompt, every model could have maximum of 8 points (since there were 8 criteria in the rubric).
The observed accuracy values were:
\\
\\
\vspace{0.5em}
\begin{tabular}{c|c}
\hline
       Gemini accuracy (rated by GPT): & 0.314 \\
       \hline
      GPT accuracy (rated by Gemini): &  0.297 \\
      \hline
\end{tabular}
\\
\\
From this, it can be deduced that, according to the secondary model, the primary target model, Gemini, has marginally higher accuracy than GPT as measured by the secondary model evaluation pipeline.

\begin{figure}[!t]
    \centering
    \includegraphics[width=\columnwidth, height=5.5cm]{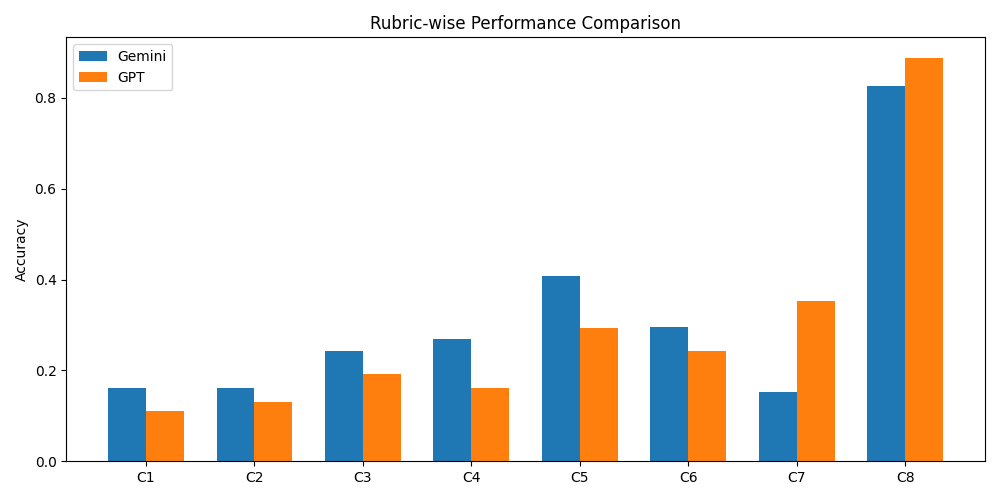}
    \caption{Comparison across rubrics.}
    \label{fig:rubriccomp}
\end{figure}

\subsubsection{\textbf{Wins and loss comparison}}
For each prompt, both target models are evaluated, and their total scores are compared. Whoever scores higher, is said to "win". The total, win, loss, and comparison are shown in  and the table below:
\\
\\
\vspace{0.5em}
\begin{tabular}{c|c}
\hline
       Gemini wins: & 37 \\
       \hline
      GPT wins: &  33 \\
      \hline
      Ties: & 29\\
      \hline
\end{tabular}
\\
\\
\subsubsection{\textbf{Accuracy across rubrics}}
For each criterion, we map the accuracy, in Figure 3. 
Referring to the criterion mapping in 
Table III, C1 captures overall functional correctness, 
C2--C4 capture structural and datapath fidelity, C5--C6 
capture sequential behavior, and C7--C8 capture visual 
quality.
From this, we can infer a few things:
\begin{itemize}

\item Gemini performs better across most semantic and functional criteria (C1--C6) when evaluated by GPT. These criteria capture the core aspects of circuit correctness, including structural fidelity, datapath integrity, control logic, and timing behavior. This suggests that Gemini is more effective at preserving the logical and functional core of the RTL description during schematic generation.

\item Even though GPT achieves higher scores on visual-oriented criteria (particularly C7 and C8), this result requires careful interpretation. During evaluation, models primarily assessed visual clarity in terms of layout cleanliness and readability, rather than strict adherence to standard schematic conventions. As a result, aspects such as the correct use of canonical symbols (e.g., AND/OR/NOR gates), standardized notation, and faithful interpretability of components were not fully enforced. Consequently, higher visual scores do not necessarily imply better semantic correctness or hardware validity.

\end{itemize}

\section{Discussion}
An important observation from our evaluation from our experiments is the near-zero value of Cohen's Kappa for both the models as highlighted in the Experimental Evaluation section V.1. This demonstrates few things:
\begin{itemize}
    \item The evaluation of LLM is not reliable. Specifically, the correctness and semantics of hardware circuit schematics cannot be correctly ascertained by LLMs, even when provided grounding by RTL codes, and validation by human judges.
    
    \item This finding raises concerns about the reliability of LLM-as-a-judge frameworks in areas requiring precise structural and semantic reasoning. While LLM-as-a-judge has shown promise in natural language tasks, our results suggest that its applicability to hardware-centric multi-modal tasks is limited.
    \item Although evaluated on hardware schematics, the proposed framework directly extends to other domains involving non-executable visual outputs, where correctness depends on structural and functional alignment for information propagation rather than exact visual equivalence.
The observed discrepancy between evaluators also reinforces the importance of human-in-the-loop validation and more rigorous evaluation strategies.

\end{itemize}
\bibliographystyle{IEEEtran}
\bibliography{references}
\end{document}